**No Sun-like dynamo on the active star ζ Andromedae from starspot asymmetry**

R. M. Roettenbacher[1], J. D. Monnier[1], H. Korhonen[2,3], A. N. Aarnio[1], F. Baron[1,4], X. Che[1], R. O. Harmon[5], Zs. Kővári[6], S. Kraus[1,7], G. H. Schaefer[8], G. Torres[9], M. Zhao[1,10], T. A. ten Brummelaar[8], J. Sturmann[8], & L. Sturmann[8]

[1]Department of Astronomy, University of Michigan, Ann Arbor, MI 48109, USA

[2]Finnish Centre for Astronomy with ESO (FINCA), University of Turku, FI-21500 Piikkiö, Finland

[3]Dark Cosmology Centre, Niels Bohr Institute, University of Copenhagen, Juliane Maries Vej 30, DK-2100 Copenhagen Ø, Denmark

[4]Department of Physics and Astronomy, Georgia State University, Atlanta, GA 30303 USA

[5]Department of Physics and Astronomy, Ohio Wesleyan University, Delaware, OH 48103 USA

[6]Konkoly Observatory, Research Center for Astronomy and Earth Sciences, Hungarian Academy of the Sciences, H-1121 Budapest, Konkoly Thege Miklós út 15-17, Hungary

[7]School of Physics, University of Exeter, Exeter, EX4 4QL, UK

[8]Center for High Angular Resolution Astronomy, Georgia State University, Mount Wilson, CA 91023, USA

[9]Harvard-Smithsonian Center for Astrophysics, 60 Garden Street, Cambridge, MA 02138, USA

[10]Department of Astronomy & Astrophysics, Pennsylvania State University, State College, PA 16802, USA

**Sunspots are cool areas caused by strong surface magnetic fields inhibiting convection[1,2]. Moreover, strong magnetic fields can alter the average atmospheric**

**structure[3], degrading our ability to measure stellar masses and ages. Stars more active than the Sun have more and stronger dark spots than in the solar case, including on the rotational pole itself[4]. Doppler imaging, which has so far produced the most detailed images of surface structures on other stars than the Sun, cannot always distinguish the hemisphere in which the starspots are located, especially in the equatorial region and if the data quality is not optimal[5]. This leads to problems in investigating the north-south distribution of starspot active latitudes (those latitudes with more spot activity), which are crucial constraints of dynamo theory. Polar spots, inferred only from Doppler tomography, could plausibly be observational artifacts, casting some doubt on their very existence[6]. Here we report imaging of the old, magnetically-active star ζ Andromedae using long-baseline infrared interferometry. In our data, a dark polar spot is seen in each of two epochs, while lower-latitude spot structures in both hemispheres do not persist between observations revealing global starspot asymmetries. The north-south symmetry of active latitudes observed on the Sun[7] is absent on ζ And, which hosts global spot patterns that cannot be produced by solar-type dynamos[8].**

ζ And is a nearby active star that is both spatially large and spotted, making it one of a small number of promising imaging targets with current interferometric capabilities. ζ And is a tidally-locked close binary (RS CVn) system consisting of a K-type cool giant and an unseen lower-mass companion star[9]. Tidal interactions have spun-up the cool primary component, causing unusually strong starspots and magnetic activity[4,10].

We observed ζ And during two observing campaigns of eleven nights spanning UT 2011 Jul 9-22 and fourteen nights spanning UT 2013 September 12-30 (see Extended Data Table 1) with the Michigan InfraRed Combiner (MIRC)[11] using all six telescopes at Georgia State University's Center for High Angular Resolution Astronomy (CHARA) Array[12] on Mount Wilson, CA, USA.

The 2011 and 2013 datasets were separately imaged onto a prolate ellipsoid via the imaging software SURFING (SURFace ImagING), an aperture synthesis imaging technique (Monnier, in preparation). This novel approach replicates the fundamental ideas behind Doppler imaging in that the whole data set is mapped onto the rotating surface at once instead of night-by-night snapshots. Treating each dataset as an ensemble also allows SURFING to fit stellar and orbital parameters (see Table 1) along with the surface temperature maps (see Figures 1 and 2).

The surface temperature maps for ζ And show peaks of 4530 K and 4550 K and minimum values of 3540 K and 3660 K in 2011 and 2013, respectively. The ~900 K range of temperatures we see across the surface is slightly larger than the ~700 K found from recent Doppler imaging work (from the Fe I 6430 Å line). A strong dark polar spot is present in both of our imaging epochs, also consistent with recent Doppler imaging studies[7,13,14]. In contrast to this persistent feature, many other large dark regions completely change from 2011 to 2013 with no apparent overall symmetry or pattern. These features and their locations are only unambiguously imaged by interferometry, since Doppler and light-curve inversion imaging techniques experience latitude degeneracies (see Methods section for more details). We now discuss the starspot implications on the dynamical large-scale magnetic field of ζ And.

The extended network of cool regions stretching across the star suggest that strong magnetic fields can suppress convection on *global* scales, not just local concentrations forming spot structures. The extent to which starspots can cover the surface of a star is presently unknown and of interest in understanding how activity saturates on rapidly-rotating, convective stars[15]. The observations in hand lend support to studies that have suggested magnetic activity can be so widespread as to alter the apparent fundamental parameters of a star[16,17]. For example, a larger region of suppressed convection gives a lower observed temperature, leading to inaccurate estimates for stellar mass and age[3]. The changes in global magnetic features will produce long-term photometric variations that are often only attributed to changes in a growing or shrinking polar starspot. We note that a polar starspot for ζ And does not affect the flux of the star as significantly as other large-scale magnetic structures due to the effects of limb darkening and foreshortening on this highly inclined system ($i$~70.0°).

The interferometric images of ζ And provide a clear confirmation of the existence of polar spots. Polar spots have been seen in Doppler images of ζ And[9,13,14] and many other active stars[4]. Polar spots produce spectral line-profile changes only in the line core itself (no Doppler shift), and the spectral signature of a symmetric polar spot is the same at each rotational phase of the star. This signature makes polar spots very easy to be produced as artifacts in the Doppler imaging process; if the depth of the spectral line-profile is not correctly modeled, then the image will exhibit a polar spot. Strong chromospheric activity has also been postulated to fill in at least some of the photospheric lines used in Doppler imaging, potentially producing a polar spot[18,19]. These facts made the reliability of polar spots highly debated in the early days of Doppler imaging[20,21] and the independent confirmation of their existence here is highly significant.

The interferometric images of ζ And presented here reveal the exact hemispheres of the spots and show strong asymmetries between the hemispheres, with 2011 map showing dominant spots on the northern and the 2013 on the southern hemisphere. On the Sun the spots are typically seen on both hemispheres in certain active mid-latitude regions, with some breaking of the symmetry in the spot numbers on the two hemispheres[7] but not as large of an asymmetry as seen on the ζ And. Such asymmetries require a departure from the solar-type $\alpha\Omega$-dynamo to a more complicated dynamo, such as one with mixed parity modes[22].

While our results only strictly apply for giant stars in RS CVn binaries, we note strong parallels between the physical conditions and magnetic behavior of these and pre-main sequence stars. To reach these conditions, the giant primary stars in RS CVn binaries rotate rapidly due to tidal spin-up and pre-main sequence stars rotate rapidly due to contraction and angular momentum transfer due to accretion of material from a circumstellar disk. These similar physical conditions hint at shared field-generation mechanisms that are observationally indistinguishable[23] and manifest as starspots. In young associations, it has been noted that derived ages are likely strongly affected by global suppression of convection[3]. These commonalities and the known consequences argue that strong stellar magnetism must be accounted for in stellar models for both pre-main sequence and giant stages of evolution for the most active stars.

Results from imaging studies using light-curve inversion and Doppler imaging techniques, as well as new interferometric spot studies[24], all re-enforce the picture that global magnetic structures cover the faces of the most active stars. Our interferometric imaging has found unambiguous signposts of these structures and clearly points to a perspective beyond the typical isolated spots observed on the Sun. The large-scale suppression of convection by these large-scale fields will have structural effects on the stellar atmosphere, including puffing up the star and decreasing the effective temperature and luminosity, dramatic alterations that must be accounted for by modern stellar structure calculations especially for young, low-mass stars that universally show strong magnetic activity[3,25]. In order to understand these structural effects, we must image with as much detail as possible more targets. The procedures used here can provide similar *H*-band images for a handful of bright, spatially-large, spotted stars (e.g. σ Geminorum, λ Andromedae). Impending advances in visible interferometry will allow for similar resolution on more stars (down to $\theta \sim 1.1$ mas). For stars that cannot be resolved in detail, combining interferometrically-observed photocenter shifts due to rotation of starspots in and out of view with Doppler imaging would resolve degeneracies inherent in the Doppler images allowing for more accurate surface maps. By acquiring a number of these maps on several stars or a few epochs of the same targets, we would be able to understand how the changing magnetic field affects our observations of stellar parameters (including mass and age)[3,26] and leads to new dynamo models, which will shed light on the impact of magnetism on stellar evolution[27,28].

**Full Methods** and associated references are available online.

**Acknowledgements**

We thank E. Pedretti for his early efforts in imaging ζ And. The CHARA Array is funded through NSF grants AST-0908253 and AST 1211129 and by Georgia State University through the College of Arts and Sciences. The MIRC instrument at the CHARA Array was funded by the University of Michigan. The 2013 CHARA/MIRC observations were supported by a Rackham Graduate Student Research Grant from the University of Michigan. This imaging work was supported by the National Science Foundation (NSF) grant AST-1311698. SURFING was partially developed with funding from the Observatoire de Paris, Meudon. This research made use of the SIMBAD database, operated at CDS, Strasbourg, France and the Jean-Marie Mariotti Center SearchCal and Aspro2 services co-developed by the FIZEAU and LAOG/IPAG. This work was supported by the Hungarian Scientific Research Fund (OTKA K-109276) and "Lendület-2009" Young Researchers' Program of the Hungarian Academy of Sciences.

**Author Contributions**

RMR, JDM, FB, XC, SK, and MZ obtained the observations of ζ And. RMR, JDM, FB, XC, SK, GHS, and MZ obtained the observations of 37 And. RMR performed the data reduction and calibration. RMR and GT determined the orbital parameters of 37 And. JDM and RMR created the images of ζ And, which were interpreted by RMR, JDM, HK, ZsK, ROH, and ANA. TAtB, GHS, JS, and LS provided observational setup and technical support.

Correspondence and requests for materials should be addressed to rmroett@umich.edu.

**Table 1 – Parameters of ζ And.**

| Parameter | Value |
|---|---|
| Angular polar diameter, $\theta_{LD}$ (mas) | 2.502 ± 0.008 |
| Polar radius ($R_\odot$) | 15.0 ± 0.8 |
| Oblateness (major to polar axis) | 1.060 ± 0.011 |
| Inclination, $i$ (°) | 70.0 ± 2.8 |
| Pole position angle (°, E of N) | 126.0 ± 1.9 |
| **Values from Literature** | |
| Distance, $d$ (pc) | 17.98 ± 0.83[29] |
| Effective temperature, $T_{eff}$ (K) | 4600 ± 100[9] |
| Luminosity, log $L/L_\odot$ | 1.98 ± 0.04[9] |
| Primary mass ($M_\odot$) | 2.6 ± 0.4[9] |
| Secondary mass ($M_\odot$) | ~0.75[9] |
| Iron metallicity [Fe/H]/[Fe/H]$_\odot$ | -0.30 ± 0.05[9] |

SURFING models assumed circular orbit ($e$ = 0) using circular radial velocity conventions with an orbital period $P_{orb}$ = 17.7694260 ± 0.00004 days and time of nodal passage $T_0$ = 49992.281 ± 0.017 (MJD)[30]. Limb darkening was held fixed with power-law exponent μ = 0.269, appropriate for ζ And based upon spectral type.

**Figure 1 – Surface image of ζ And from 2011 July with eleven nights of data using SURFING.** **a**, The temperature of ζ And is presented in an Aitoff projection. The contours represent every 200 K from 3400-4600 K. The dashed line at the bottom pole is hidden due to inclination. Arrows point to example starspots. **b**, The surface reflects how the star is observed on the sky with $H$-band intensities (mean $H = 1.64$). The $\phi = 0.000$ plot shows longitude 0° at the bottom right of the star with 90° across the middle. The phases assume circular orbit radial velocity conventions.

**Figure 2 – Surface images of ζ And from 2013 September with fourteen nights of data using SURFING.** These images are presented as in Figure 1, except the 200 K contours of the Aitoff projection (**a**) range from 3600-4600 K. The polar spot is observed to have evolved between the two sets of observations. The lower-latitudes spot present in the 2011 data set are not present in the 2013 data set, with the new spots mostly located below the equator, emphasizing the spot-latitude asymmetry observed.

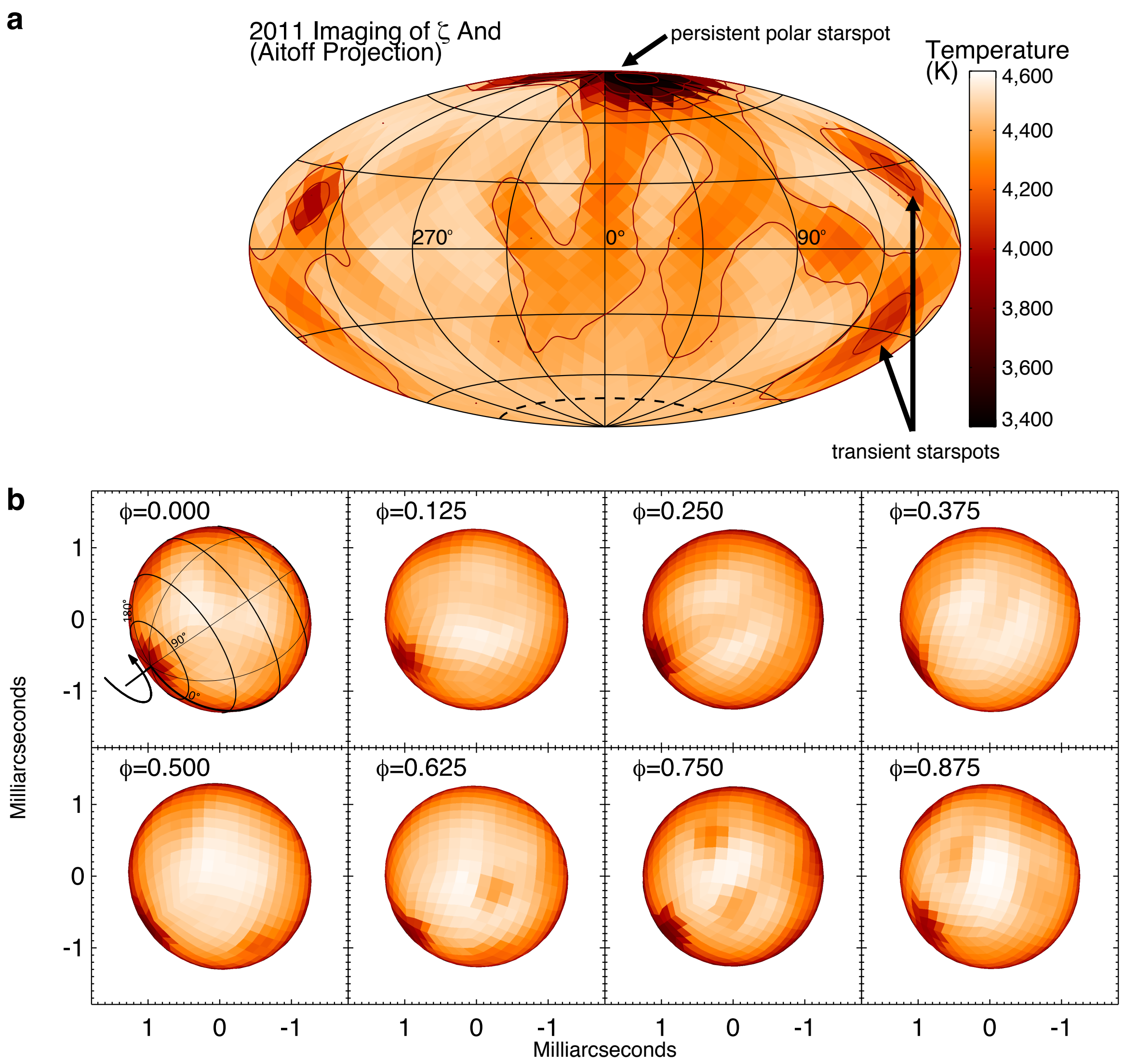

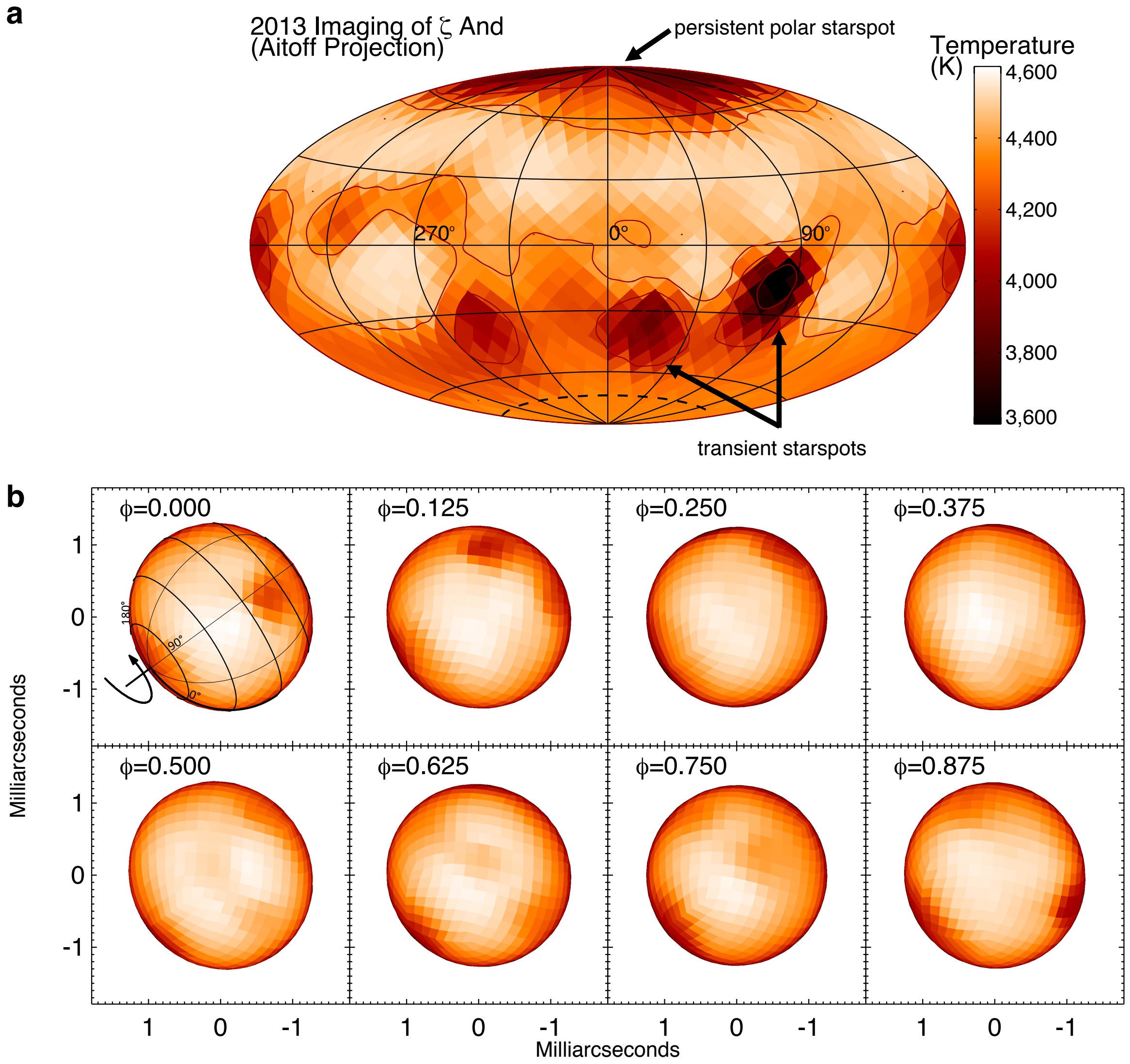

## Methods

### Starspot Imaging Methods

Large spots covering a substantial fraction of the stellar surface have been indirectly imaged using light-curve inversion and Doppler imaging techniques[31,32]. Light-curve inversion reproduces spotted stellar surfaces based on time-series data and can only reproduce structures observed as rotational modulations—structures such as static polar spots are practically invisible to light-curve inversion imaging. Light-curve inversion typically reveals only weak relative latitude information for the spots, although this can be improved with the combination of concurrent observations in multiple filters[33]. A more detailed surface map, both in latitude and longitude, can be obtained with Doppler imaging, which creates surface temperature maps from tracking small changes in absorption lines as starspots rotate in and out of view[32]. Still, this method cannot always distinguish the hemisphere in which the structures are located. In order to confirm the important findings from these methods and to firmly understand global characteristics of activity that can alter stellar radii and effective temperature, a more direct imaging method is required that is immune to these ambiguities.

The nearest magnetically-active stars are too small to be resolved by even our largest telescopes. However, long-baseline interferometry has the potential to image sub-milliarcsecond features on the surfaces of nearby stars. To date, interferometric imaging has been successfully used to confirm the oblateness and gravity darkening of rapidly-rotating stars[34] and even to image a spotted stellar surface[24]. To improve the resolution and imaging quality for rotating stars that show strong magnetic activity, we debut here an "imaging-on-a-sphere" technique that uses interferometeric observations from multiple nights to constrain a surface temperature map.

This naturally takes advantage of the multiple views we have of starspot structures as they rotate across the disk of the star. Interferometric imaging produces unique images that resolve the degeneracies in latitude of Doppler imaging and provides the first independent confirmation of the existence of polar starspots.

## Interferometric Data

Our interferometric observations were obtained using the Michigan InfraRed Combiner (MIRC) at the Center for High Angular Resolution Astronomy (CHARA) Array. The CHARA Array consists of six 1-m telescopes in a Y-shaped configuration with baselines ranging from 34 to 331 m[12]. For the ζ And data, MIRC[11] combined light from all six CHARA Array telescopes in the *H*-band (eight channels across 1.5-1.8 μm for $\lambda/\Delta\lambda \sim 40$), resulting in an angular resolution of $\lambda/2B \sim 0.5$ milliarcseconds.

The data were reduced and calibrated with the standard MIRC pipeline[35]. We searched without success for evidence of the faint companion in our interferometry data using our proven grid search method[36], and could only secure a 1-σ lower limit of 300:1 on the *H*-band flux ratio between primary and companion.

The data products obtained from reducing the CHARA/MIRC data with the standard pipelines consist of visibilities, closure phases, and triple amplitudes. Representative samples of the visibilities and closure phases are presented in Extend Data Figures 1 and 2 for a single night (UT 2013 September 15) of six-telescope CHARA/MIRC observations of ζ And. The reduced data are available in OI-FITS format[37] upon request.

## Calibration Stars

The twenty-five nights of interferometric data span 2011 July 9-22 and 2013 September 12-30. For these nights of observation we use four calibration stars (37 Andromedae, γ Pegasi, γ

Trianguli, and 58 Ophiuchi) interspersed with observations of the target star ζ And. γ Peg, γ Tri, and 58 Oph are modeled as spherical, uniform disk stars[38] with their parameters included in Extended Data Table 2.

The calibrator 37 And is a recently-discovered binary system[39] with primary-to-secondary *H*-band flux ratio of 80 ± 20. Ordinarily, binary stars make poor calibrators, but the 37 And system was observed enough times to determine its orbit precisely and salvage its use for calibrating our primary target ζ And. We detect the companion of 37 And in nineteen nights of data (see Extended Data Table 3) using a grid search for the companion. To constrain orbital parameters, we combined the visual orbit with the primary star's radial velocity curve obtained with archival spectra from the ELODIE high-resolution échelle spectrograph formerly on a 1.93-m telescope at Observatoire de Haute-Provence, France[40]. Extended Data Figures 3 and 4 show the system orbit and radial velocity curve and Extended Data Table 4 contains the system orbital parameters. The orbital parameters are used in the MIRC calibration pipeline to account for the effect of the companion of 37 And.

## SURFING imaging code

The image reconstruction code SURFING (SURFace imagING) was specially written for this project: to image surfaces of rotating stars. We create a global model of the star, including geometrical parameters (polar radius, oblateness, inclination, pole position angle, limb darkening coefficient, rotational period, epoch) as well as the surface temperature map. We cover the surface with tiles of equal area using the HEALPix methodology[41], using 768 tiles to match the spatial resolution of CHARA. Each tile has an area of 0.025 $mas^2$. We represent the shape of the stellar photosphere as a prolate spheroid to approximate a slightly partially-filling Roche potential. The deviation from spherical appears as gravity darkening and accounts for a temperature difference of only ~60 K, which is much smaller than the temperature variations of the starspots.

We sampled the large range of geometrical models using an affine invariant ensemble Markov chain Monte Carlo approach[42], with a nested loop to iteratively optimize the surface temperature map within each walker of the outer loop (this was needed as the 768 free parameters needed to characterize the temperature map would be intractable using a Markov chain approach). Extensive testing was carried out using blind simulated data to optimize the speed of convergence. In addition to minimizing the $\chi^2$ statistic, we also incorporated priors on each parameter and could experiment with a variety of imaging regularizers, such at Total Variation (TV) or L2norm of wavelet coefficients.

In addition to testing the code on simulated data, we were also able to check results using soon-to-be-published data on the spotted star λ And[24], finding comparable results and confirming the inclination and position angle of the pole. We also checked that the imaging-derived orientation of another RS CVn (σ Gem, Roettenbacher et al. in preparation) matched the orbital plane of the close companion. The results of these tests and additional details about the implementation of our method will be described in a future paper (Monnier et al., in preparation).

**ζ And Parameters from SURFING**

Another test of the robustness of our fitting methodology is to determine the stellar orientation for independent data sets and compare the results. While the magnetic field structures will vary from year to year, the inclination of the pole and its position angle on the sky will not. Extended Figure 5 shows a $\chi^2$ surface for three separate years: a 2008 pilot set of observations using the MIRC four-telescope system, the 2011 dataset, and the 2013 dataset. This figure shows two large regions of reduced $\chi^2$ around inclination $i$ = 90° and PA = -60° or 120°. These regions reflect the oblateness of the star and the basic orientation on the sky. The region near inclination $i = 70°$, PA = 120° has the lowest $\chi^2$ in all years, especially in 2013. This tells us the spots move from south-west toward north-east and not the other way around, and the consistent picture from year to year both shows the efficacy of the code and gives confidence that we are measuring true astrophysical signal and not over-fitting noise or systematic errors.

The results from these grid studies have been used to robustly estimate the geometrical parameters for ζ And, and these parameters are found in Table 1. The error bars associated with the parameters were determined by combining the results of data sets from 2011 and 2013, with the additional data from 2008. As the 2008 data were obtained with only four CHARA telescopes, they are not of high enough quality for imaging.

To convert *H*-band intensities from the reconstructed images into photospheric temperatures, we utilized Kurucz atmospheric models[43] for [Fe/H]=-0.25 and appropriate log *g*. Note that the overall temperature scale in our maps is uncertain (overall multiplicative scaling) due to lack of coeval photometry at *H* band; here we adopted mean *H*-band magnitude of $H = 1.64$ based on archival infrared photometry.

**ζ And Imaging Tests**

The previous section laid out our robust method for determining the geometrical parameters of ζ And by using three independent observing data from different years. The next issue is to determine the reliability of the surface temperature maps. We split the extensive CHARA/MIRC data from 2013 into two sets of seven nights each, alternating chronologically which night went to which set. Since spots take many days to cross the face of the star, each dataset should be viewing the same spots even though the timing was different. We present a comparison of the SURFING results in Extended Figure 6. Both partial datasets reproduce the main features observed by the full dataset shown in the main text. This proves that the features seen in the maps are real and are not the result of an over-fitting to poor quality data or peculiarities of the nightly *uv* coverage.

The unprecedented phase coverage in the 2011 and (especially) the 2013 observing runs have allowed for textbook imaging fidelity tests for ζ And. We show the large-scale dark regions that cover ζ And are highly robust and statistically significant, likely deriving from magnetically-suppressed convection.

**Code Availability**

At present, we have opted not to make the SURFING code available because of a publication in preparation, which will detail the use and the applicability of the resource.

Methods References

**Extended Data Table 1 – Observations and calibrators of ζ And.**

**Extended Data Table 2 – Uniform disk sizes (*H*-band) of calibrators.**

The uniform disk diameters were obtained with SearchCal[38].

**Extended Data Table 3 – Binary separation and position angle measurements of 37 And.**

**Extended Data Table 4 – Orbital parameters of 37 And.**

The orbital parameters were obtained by combining the ELODIE radial velocity curve with the CHARA/MIRC detections.

| UT Date | Modified Julian Date (MJD) | Calibrators Used |
|---|---|---|
| 2011 Jul 9 | 55751.536 | 37 And |
| 2011 Jul 10 | 55752.531 | γ Peg |
| 2011 Jul 11 | 55753.480 | 37 And |
| 2011 Jul 12 | 55754.469 | γ Peg |
| 2011 Jul 14 | 55756.478 | γ Peg |
| 2011 Jul 16 | 55758.505 | 58 Oph |
| 2011 Jul 17 | 55759.481 | γ Peg |
| 2011 Jul 19 | 55761.475 | 37 And, γ Peg |
| 2011 Jul 20 | 55762.517 | γ Peg |
| 2011 Jul 21 | 55763.478 | γ Peg, γ Tri |
| 2011 Jul 22 | 55764.480 | γ Peg, γ Tri |
| 2013 Sep 12 | 56547.449 | 37 And, γ Tri |
| 2013 Sep 13 | 56548.426 | 37 And, γ Tri |
| 2013 Sep 15 | 56550.392 | 37 And, γ Peg, γ Tri |
| 2013 Sep 16 | 56551.365 | 37 And, γ Peg, γ Tri |
| 2013 Sep 17 | 56552.345 | 37 And, γ Tri |
| 2013 Sep 18 | 56553.359 | 37 And, γ Peg, γ Tri |
| 2013 Sep 19 | 56554.407 | 37 And, γ Peg, γ Tri |
| 2013 Sep 20 | 56555.365 | 37 And, γ Peg, γ Tri |
| 2013 Sep 21 | 56556.403 | 37 And, γ Tri |
| 2013 Sep 23 | 56558.403 | 37 And, γ Tri |
| 2013 Sep 24 | 56559.343 | 37 And, γ Peg, γ Tri |
| 2013 Sep 28 | 56563.367 | 37 And, γ Tri |
| 2013 Sep 29 | 56564.357 | 37 And, γ Tri |
| 2013 Sep 30 | 56565.334 | 37 And, γ Tri |

| **Star Name (HD number)** | **$\theta_{UD}$ (mas)** |
| --- | --- |
| γ Peg (HD 886) | 0.41 ± 0.03 |
| γ Tri (HD 14055) | 0.51 ± 0.03 |
| 58 Oph (HD 160915) | 0.68 ± 0.05 |

| UT Date | Modified Julian Date (MJD) | Separation (mas) | Position Angle (°, E of N) | Error Ellipse Major Axis (mas) | Error Ellipse Minor Axis (mas) | Error Ellipse Position Angle (°) |
|---|---|---|---|---|---|---|
| 2011 Jul 9 | 55751.492 | 11.46 | 217.0 | 0.13 | 0.11 | 300 |
| 2011 Jul 12 | 55754.499 | 9.09 | 230.6 | 0.10 | 0.08 | 270 |
| 2011 Jul 19 | 55761.465 | 6.38 | 295.4 | 0.06 | 0.05 | 330 |
| 2011 Nov 29 | 55894.159 | 63.57 | 143.9 | 1.82 | 0.16 | 340 |
| 2013 Sep 12 | 56547.470 | 82.28 | 152.6 | 0.58 | 0.15 | 330 |
| 2013 Sep 13 | 56548.399 | 82.51 | 152.7 | 0.02 | 0.10 | 30 |
| 2013 Sep 15 | 56550.340 | 82.47 | 152.8 | 0.81 | 0.26 | 350 |
| 2013 Sep 16 | 56551.344 | 82.73 | 152.9 | 1.23 | 0.27 | 340 |
| 2013 Sep 17 | 56552.329 | 82.63 | 152.9 | 0.48 | 0.17 | 330 |
| 2013 Sep 18 | 56553.250 | 82.88 | 153.0 | 0.95 | 0.21 | 340 |
| 2013 Sep 19 | 56554.292 | 83.06 | 153.1 | 0.72 | 0.27 | 350 |
| 2013 Sep 20 | 56555.301 | 83.02 | 153.2 | 0.79 | 0.14 | 330 |
| 2013 Sep 23 | 56558.416 | 83.08 | 153.4 | 1.01 | 0.22 | 340 |
| 2013 Sep 24 | 56559.239 | 83.29 | 153.5 | 0.58 | 0.23 | 340 |
| 2013 Sep 28 | 56563.303 | 83.62 | 153.7 | 1.08 | 0.21 | 340 |
| 2013 Sep 30 | 56565.276 | 83.65 | 153.8 | 0.83 | 0.20 | 340 |
| 2014 Aug 19 | 56888.422 | 14.45 | 98.2 | 0.32 | 0.18 | 280 |
| 2014 Aug 20 | 56889.446 | 15.58 | 100.8 | 0.24 | 0.17 | 10 |
| 2014 Sep 1 | 56901.412 | 22.82 | 116.9 | 0.11 | 0.07 | 300 |

| **Parameter** | **Value** |
|---|---|
| Semimajor axis, $a$ (mas) | 46.66 ± 0.06 |
| Eccentricity, $e$ | 0.8405 ± 0.0009 |
| Inclination, $i$ (°) | 52.5 ± 0.3 |
| Argument of periastron, ω (°) | 168.9 ± 0.3 |
| Ascending node, Ω (°) | -17.6 ± 0.2 |
| Orbital period, $P_{orb}$ (days) | 550.7 ± 0.2 |
| Time of periastron passage, $T_0$ (MJD) | 55765.45 ± 0.04 |
| Velocity semi-amplitude, $K_A$ (km/s) | 11.1 ± 0.5 |
| System velocity, γ (km/s) | 5.33 ± 0.07 |

**Extended Data Figure 1 – Visibility curve of UT 2013 September 15 observations of ζ And with CHARA/MIRC.** The observed visibilities are plotted in black with 1σ error bars and the SURFING model visibilities are overlaid in red.

**Extended Data Figure 2 – Closure phases of UT 2013 September 15 observations of ζ And with CHARA/MIRC.** Each block represents a temporal block of observations with data plotted in black (with 1σ errors) and SURFING model in red.

**Extended Data Figure 3 – Orbit of 37 And.** The gray plus signs represent measurements of the companion (1σ errors on detections are smaller than the symbols). The observed resolved disk of 37 And is plotted as the black dot at the origin. The thin solid black line is the best-fit orbit from combining the interferometric detections and the ELODIE radial velocities. Note: the axis units are milliarcseconds (mas) with north up and east to the left.

**Extended Data Figure 4 – Radial velocity curve of the primary star of 37 And.** The data points are based upon archival ELODIE spectra (with 1σ error bars). The orbital solution used the velocity measurements and the interferometric measurements simultaneously. The solid line is the best-fit orbit and the gray lines are fifty Monte Carlo realizations of the orbit.

**Extended Data Figure 5 – Reduced, normalized $\chi^2$ surface plot of the SURFING model fit for position angle and inclination of ζ And.** The peak is found at PA = 126.0° ± 1.9 and $i$ = 70.0° ± 2.8. The epoch of each data set is located in the upper left of the panel.

**Extended Data Figure 6 – Comparison of the 2013 CHARA/MIRC data set divided into two sets of seven nights of data.** These projections are plotted as the Aitoff projection in Figure 2.

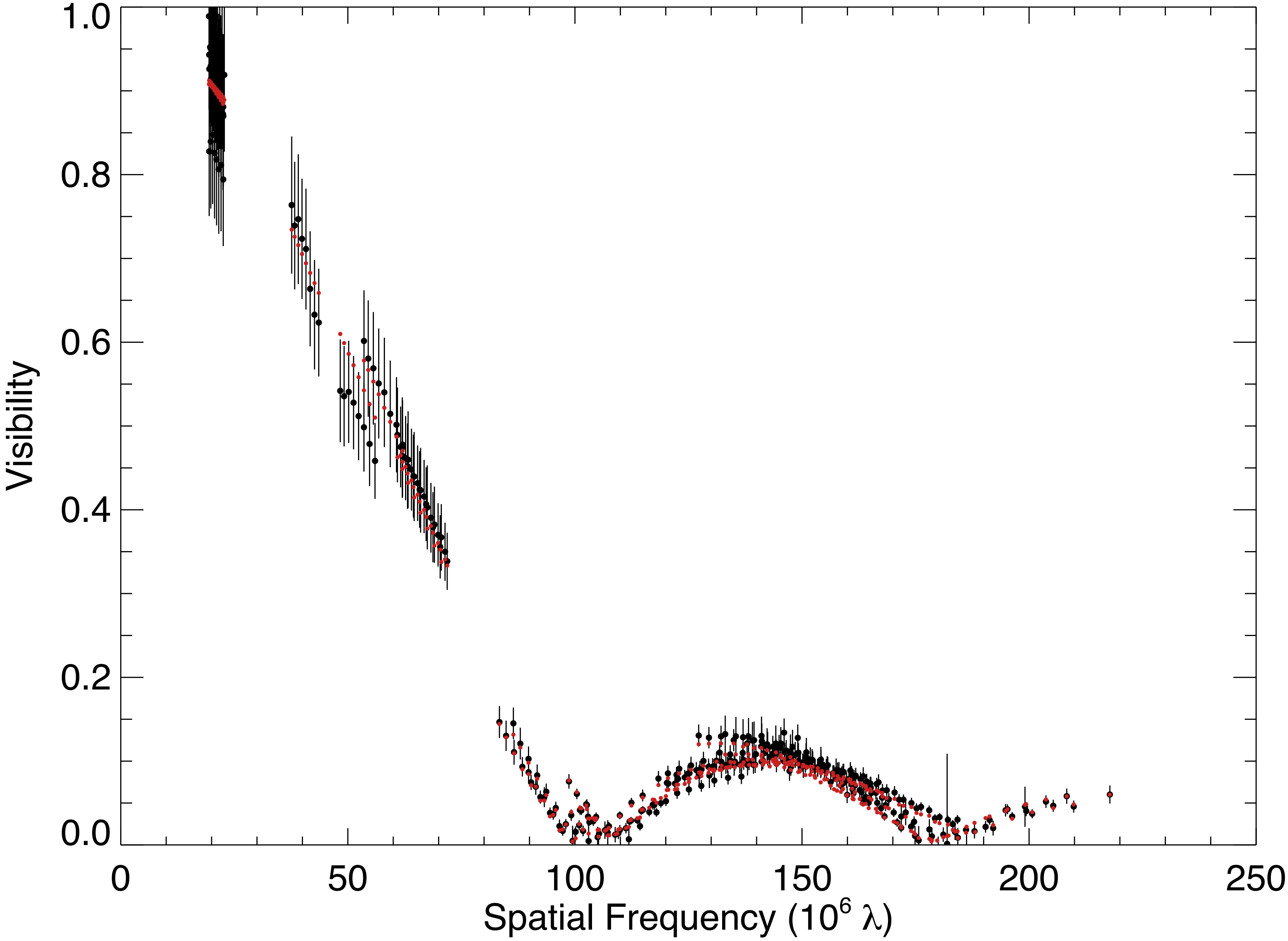

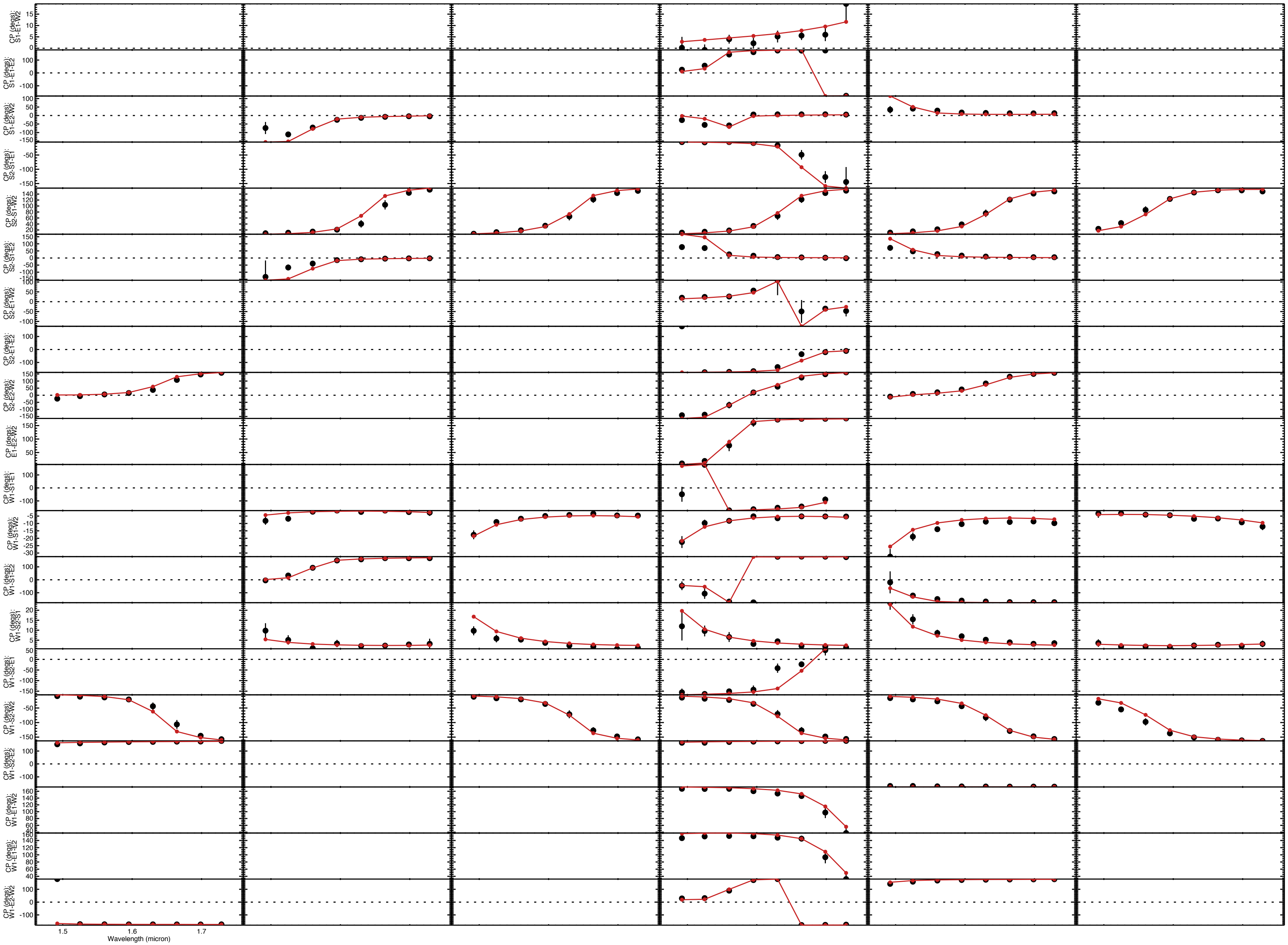

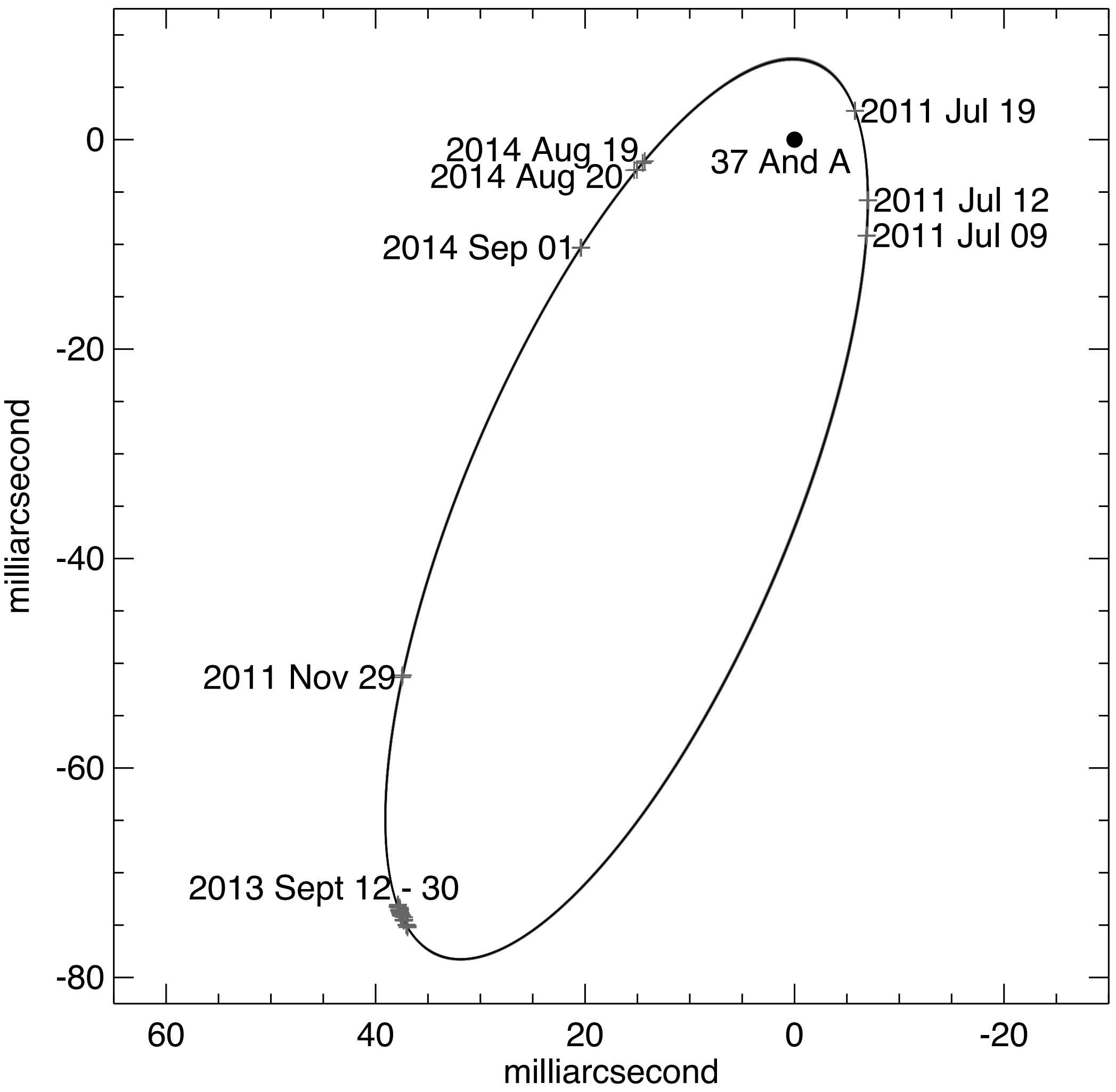

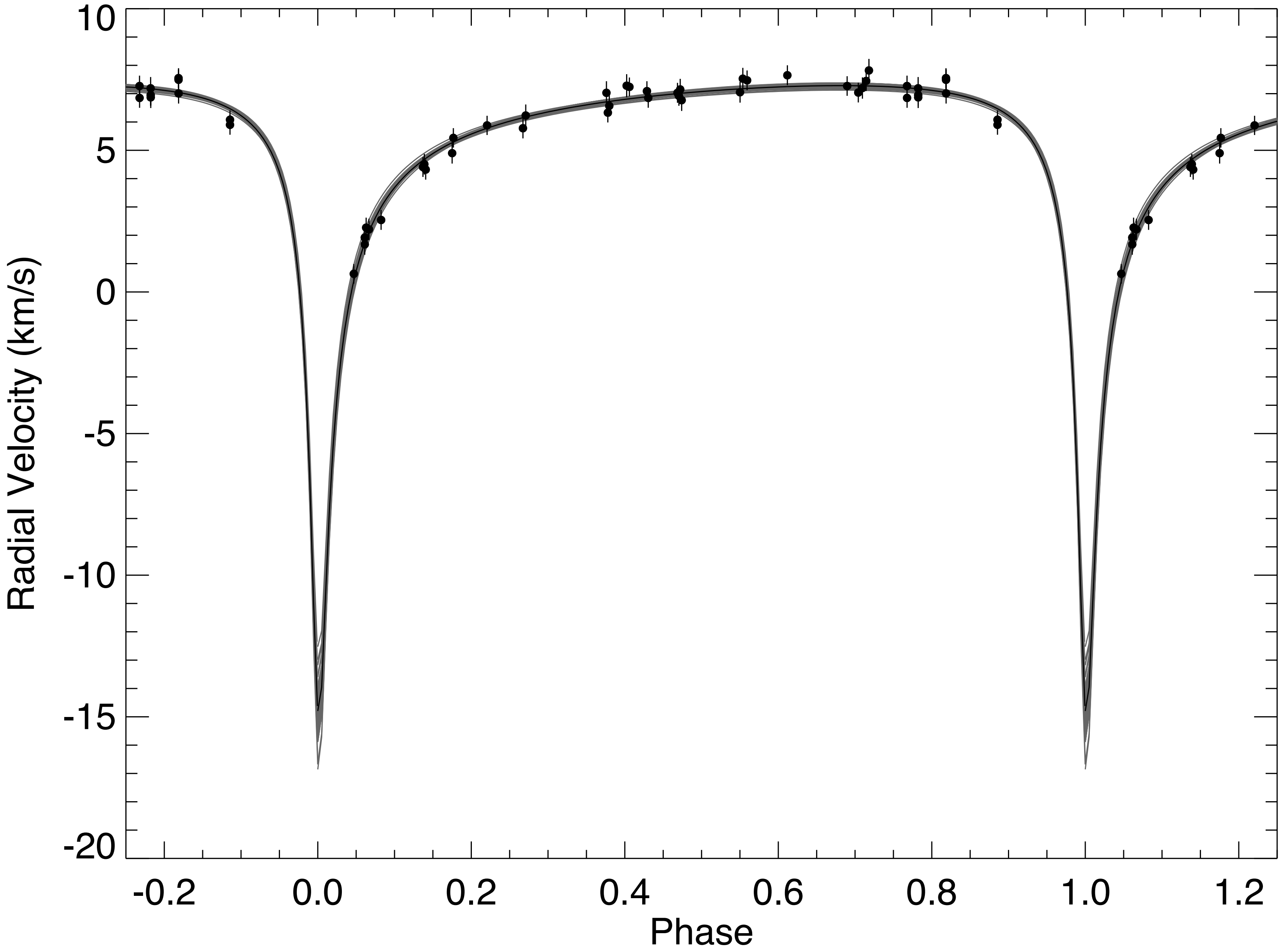

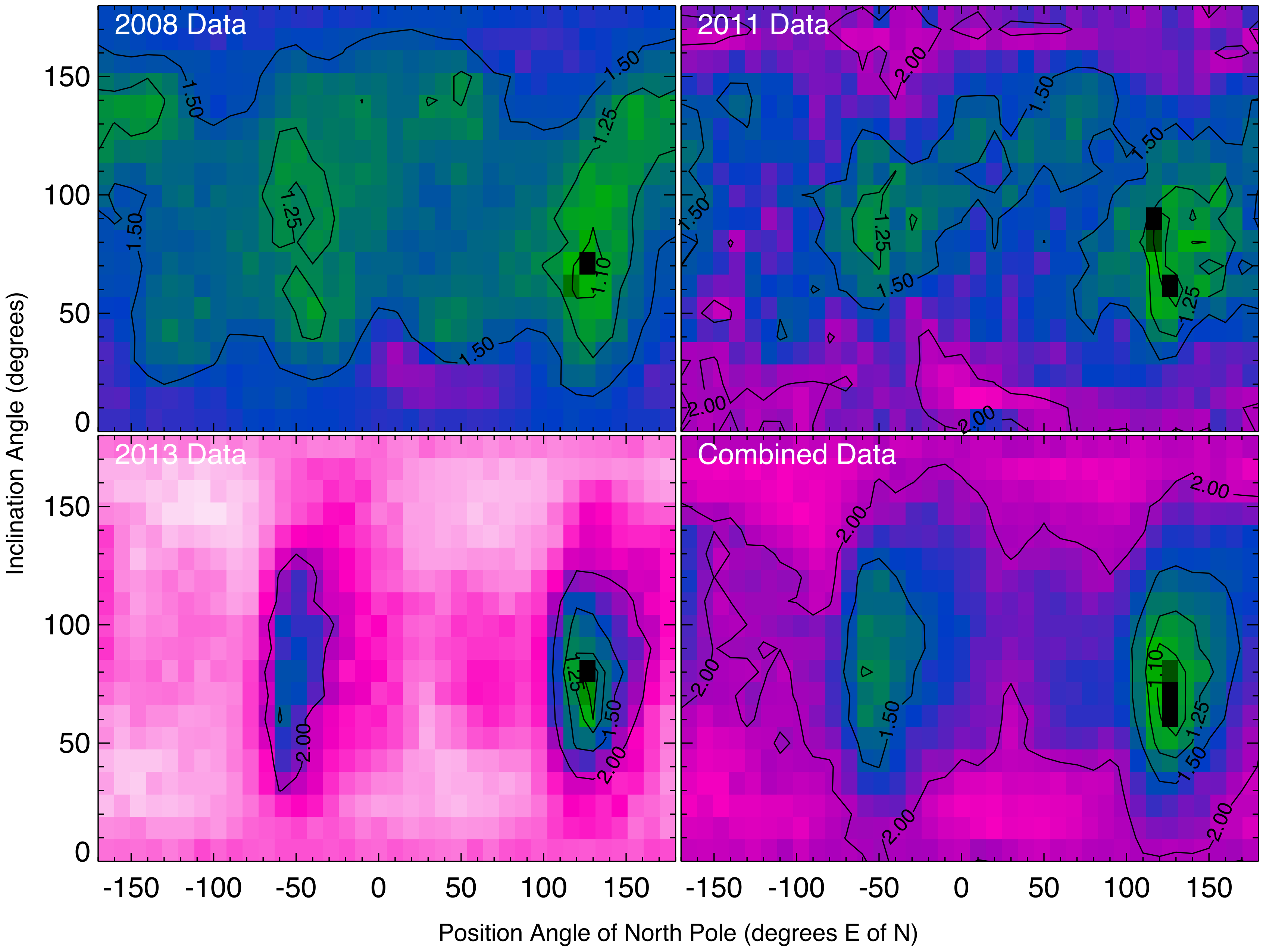

2008 Data
2011 Data
2013 Data
Combined Data
Inclination Angle (degrees)
Position Angle of North Pole (degrees E of N)

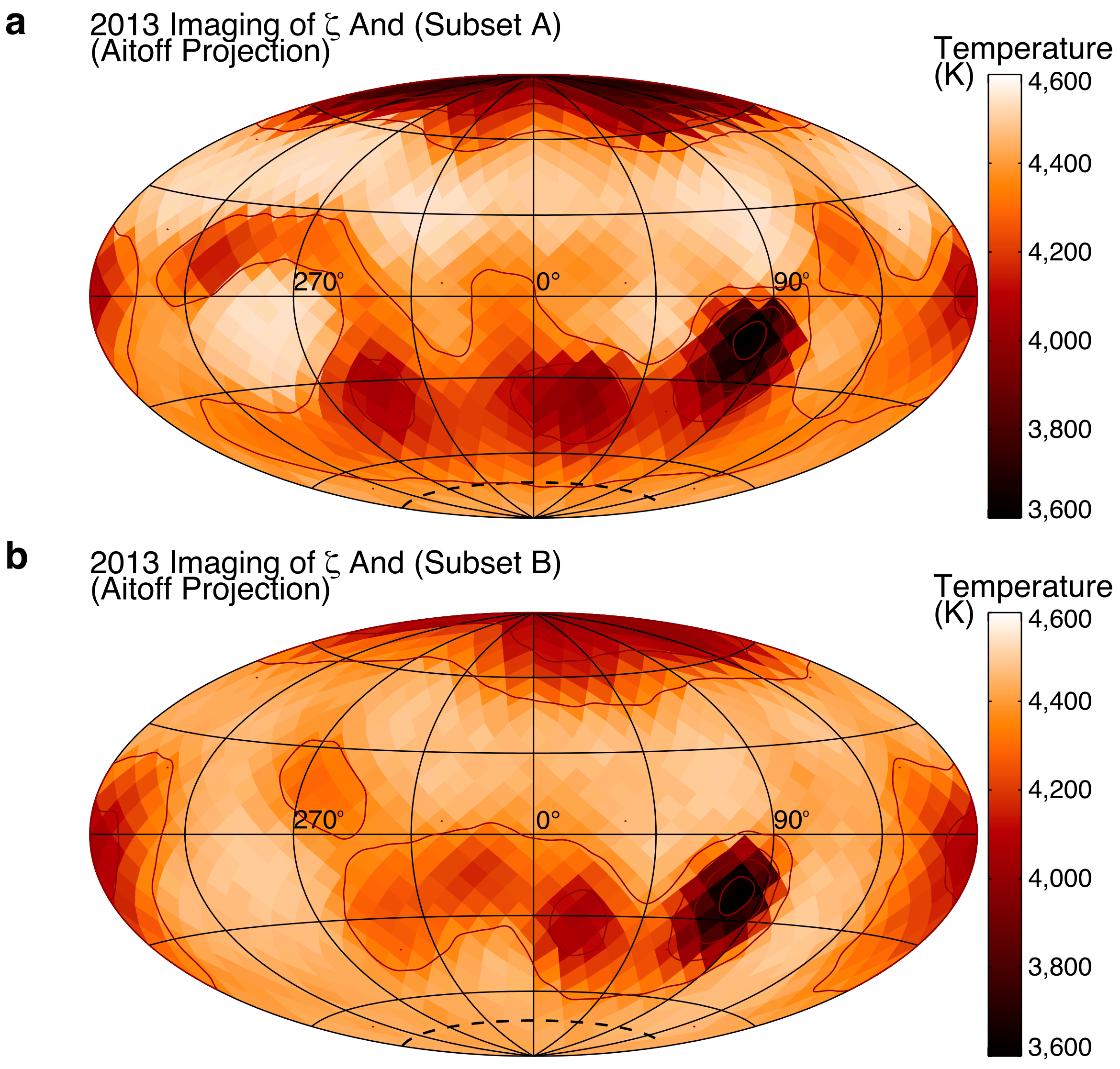